# Magnetic-Flux-Flow Diagrams for Design and Analysis of Josephson Junction Circuits

Yongliang Wang

*Abstract*—Josephson junction circuits, such as superconducting quantum interference devices (SQUIDs) and single-flux-quantum (SFQ) circuits, have been applied in both analog and digital systems for their ultralow-noise, high-speed, and power-efficient features. However, their analyses are not well supported by the conventional circuit diagrams and the charge-based analysis methods, due to the Josephson effects and flux quantization phenomena. This article introduces a type of magnetic-flux-flow (MFF) diagrams for the design and analysis of Josephson junction circuits. The MFF diagram of a Josephson junction circuit is the graphical expression of circuit equations. It not only gives the behaviors of Josephson junctions and superconducting loops new physical meanings, but also visualizes how the flux quanta are transferred among superconducting loops. The advantages of MFF diagrams are demonstrated in the design and analysis of typical Josephson junction circuits. It is shown that the MFF diagram and flux-based analysis method are the indispensable complement to the conventional circuit methodologies for Josephson junction circuits.

## I. Introduction

JOSEPHSON junctions have create various ultra-low noise, power-efficient and high-speed superconducting circuits [1] applied for both analog and digital domains, being complementary to semiconductor electronics. Superconducting quantum interference devices (SQUIDs) are the first type of practical Josephson junction circuits [2]; they are the ultra-sensitive flux-to-voltage converters widely used in magnetic field measurements [3, 4]. Single flux quantum (SFQ) circuits are ultrahigh-speed and low-power Josephson junction digital circuits [5, 6]; they are promising for the next generation signal processing and computing systems [7, 8]. Moreover, Josephson junctions have also been applied to develop quantum electromagnetic circuits for quantum computing [9, 10].

Josephson junction circuits are the networks of Josephson junctions and superconducting wires, as illustrated in Fig.1. They are distinct from normal resistor-inductor-capacitor (RLC) circuits for the unique Josephson effects [12] and flux quantization phenomena [13]. In the circuits shown in Fig. 1, any node-voltage $v_n$ ($n =1, 2...$) can be redefined with the macroscopic nodal phases $\varphi_n$, namely $d\varphi_n/dt = 2\pi v_n/\Phi_0$. For the design and analysis of Josephson junction circuits, it is not difficult to transform the voltage-based nodal analysis method [14] into the phase-based nodal analysis method [15], based on which the SPICE (simulation program with integrated circuit emphasis) tools, such as JSIM [16], JSPICE [17], PSCAN [18], and JoSIM [19], are developed. Those tools can calculate the voltage and current responses of all the Josephson junctions in a given Josephson junction circuit, as long as we can provide the conventional circuit diagram.

However, we find that the conventional circuit diagram and the nodal analysis method are inconvenient for the design and analysis of Josephson junction circuits. First, conventional circuit diagrams have deconstructed circuit loops into branches and internalized the magnetic-flux effects with inductors and transformers; they are difficult to describe the flux-quantization phenomena, such as the flux-trapping, in superconducting loops, considering the mutual couplings between loops. Second, the '0' and '1' states of SFQ logics cannot be directly read from the node phases or voltage pulses of Josephson junctions, and are difficult to understand for the electronic engineers, who are trained with the complementary metal oxide semiconductor (CMOS) logics and transistor-transistor logics (TTL), which '0' and '1' logic states are simply read from the node voltages according to voltage levels.

Those problems can be explained by the charge-flux duality of electric circuits. An electric circuit is both an electric-charge distribution and a magnetic-flux distribution networks [20], as illustrated in Fig. 2. The circuit viewed from nodes is an electric-charge distribution network, where electric charges are flowing from node to node; viewed from loops, it is also a magnetic-flux distribution network, where the magnetic fluxes generated by loop currents are transferred from loop to loop. CMOS circuits, which express logic signal with node voltages, are the electric-charge distribution networks, while Josephson junction circuits, which achieve flux-sensitive characteristics due to the flux quantization effects, are better to be treated as the magnetic-flux distribution circuits. Conventional circuit diagrams consisted of branches and nodes exhibit the paths of electric-charge flows; the nodal analysis method depict the distributions of electric charges through solving the nodal voltages; they are apparently suitable for CMOS circuits rather than Josephson junction circuits. Josephson junction circuits call for flux-based circuit diagrams and analysis methods.

In this article, we introduce a type of magnetic-flux-flow (MFF) diagrams to implement the flux-based analysis method for Josephson junction circuits. The MFF diagram of an electric circuit is the projection of the circuit viewed from magnetic fields, and thereby can be directly transformed to the circuit diagram for physical implementations. They intuitively depict

Yongliang Wang is with the State Key Laboratory of Functional Materials for Informatics, Shanghai Institute of Microsystem and Information Technology, Chinese Academy of Sciences (CAS), and the CAS Center for Excellence in Superconducting Electronics (CENSE), Shanghai 200050, China (e-mail: wangyl@mail.sim.ac.cn).

the dynamic of magnetic-flux distributions inside the Josephson junction circuits, and can be directly simulated with the general circuit equations. We demonstrated the advantages of MFF diagrams in the analyses of typical Josephson junction circuits. It is shown that the MFF diagrams of SFQ circuits are as readable and understandable as CMOS logic circuits. For example, from the MFF diagram of SFQ circuits, we can clearly see that the bits for SFQ logics are represented by the flux quanta coupled in loops, and Josephson junctions toggle the SFQ bits similar as diodes rectify the node voltages in semiconductor circuits.

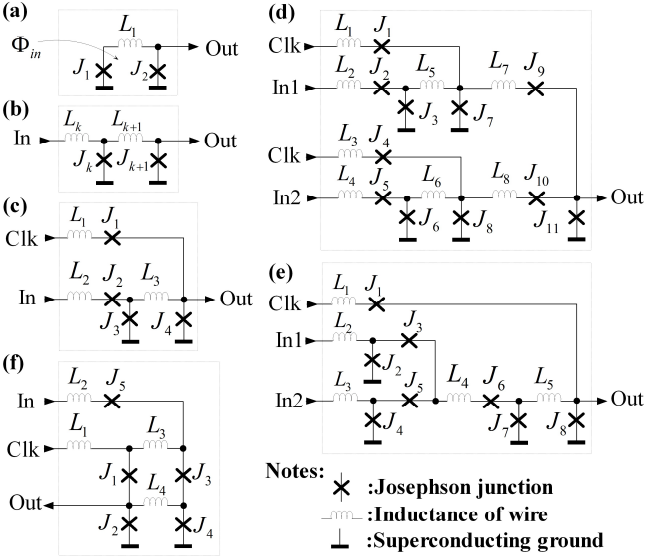

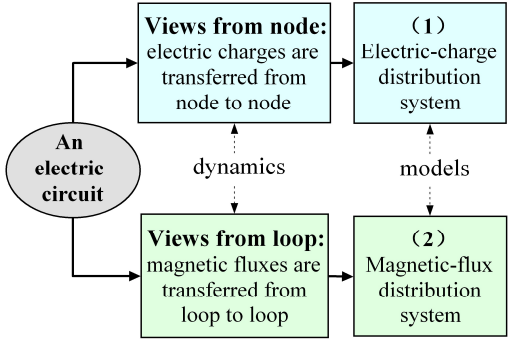

Fig. 1. Typical Josephson junction circuits [11]: (a) direct current (dc) SQUID; (b) Segment of the Josephson transmission line (JTL); (c) D-type flip-flop (DFF) circuit; (d) AND-gate circuit; (e) OR-gate circuit; (f) NOT-gate circuit.

Fig. 2. Flux-charge duality of electric circuits: A circuit viewed from nodes is an electric-charge distribution system; it is also a magnetic-flux distribution system, if it is viewed from loops.

## II. THEORY

### A. Two Objects in Josephson Junction Circuits

From the magnetic-field point of view, an electric circuit is a magnetic-flux distribution network, in which loops are circulating with eddy currents and are magnetically coupled. Based on this understanding, we can deconstruct any Josephson junction circuit into a group of independent superconducting loops with Josephson junctions embedded [20].

Assuming that there are $P$ loops and $Q$ Josephson junctions ($P$, $Q$ are integers) in a given Josephson junction circuit, as shown Fig. 3(a) and (b), we define each loop, namely Loop-$i$, with a loop-current $i_{mi}$, an external flux $\Phi_{ei}$, and the total flux coupled in the loop $\Phi_{mi}$.

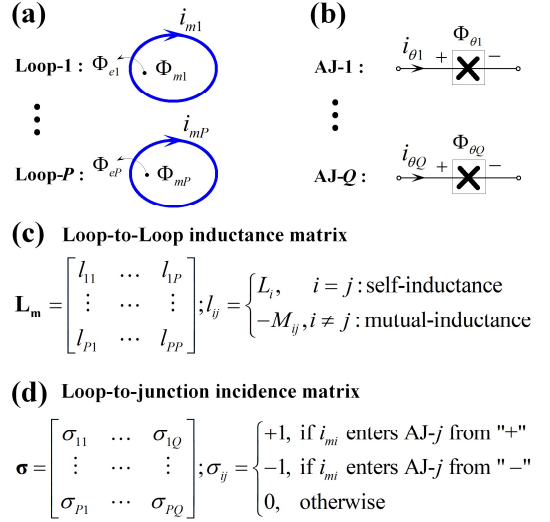

Fig. 3. Components and topology information for a given Josephson junction circuit: (a) superconducting loops; each loop, Loop-$i$ ($i$ =1, …, $P$), contains a coupled flux $\Phi_{mi}$ and a loop-current $i_{mi}$ induced by an external flux $\Phi_{mi}$. (b) Current-biased Josephson junctions, each junction, AJ-$j$ ($j$ =1, …, $Q$) is defined with a branch current $i_{\theta j}$ and a nominal flux output $\Phi_{\theta j}$. (c) Loop-to-loop inductance matrix, where $L_i$ is the self-inductance of Loop-$i$, and $M_{ij}$ is the mutual-inductance between Loop-$i$ and Loop-$j$; $L_i$ is always positive; $M_{ij}$ is positive only if Loop-$j$ with positive $i_{mj}$ generates negative fluxes in Loop-$i$. (d) Loop-to-junction incidence matrix; the element $\sigma_{ij}$ describes whether AJ-$j$ is embedded in Loop-$i$ and in what polarity it is inserted.

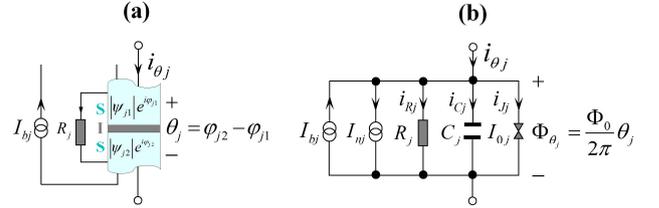

Fig. 4. Current-biased Josephson junction and its equivalent circuit: (a) structure of AJ-$j$, where, $\varphi_{j1}$ and $\varphi_{j2}$ are the quantum phases of the macroscopic wave functions of two superconductors; $R_j$ is the shunt resistance; $I_{bj}$ is the bias current. (b) Equivalent circuit of AJ-$j$ driven by the branch current $i_{\theta j}$, where $\Phi_{\theta j}$ is the nominal flux output at two terminals; $I_{0j}$ is the critical current of Josephson junction and $C_j$ is the junction capacitance. The input current from $I_{bj}$ and $i_{\theta j}$ are converted into the Josephson current $i_{Jj}$, displacement current $i_{Cj}$, normal current $i_{Rj}$, and noise current $I_{nj}$.

The magnetic fluxes coupled in those superconducting loops are expressed as

$$\mathbf{\Phi_m} = \mathbf{L_m} \cdot \mathbf{i_m} - \mathbf{\Phi_e} \qquad (1)$$

where the vectors $\mathbf{\Phi_m} = [\Phi_{m1}, …, \Phi_{mP}]^T$, $\mathbf{\Phi_e} = [\Phi_{e1}, …, \Phi_{eP}]^T$, $\mathbf{i_m} = [i_{m1}, …, i_{mP}]^T$, and $\mathbf{L_m}$ is a loop-to-loop inductance matrix introduced to describe the self and mutual inductances of those magnetically-coupled loops, as illustrated in Fig. 3(c).

Each Josephson junction, namely AJ-$j$ ($j$ =1, 2, …, $Q$), is an active junction (AJ) biased with the external current source, as depicted in Fig. 4(a). AJ-$j$ generates a phase difference $\theta_j$ at two terminals, when it is driven by the bias current $I_{bj}$ and the branch current $i_{\theta j}$. Its equivalent circuit is drawn as shown in Fig. 4(d), according to the resistively-capacitively-shunted junction (RCSJ) model [21]. This AJ-$j$ would work as a current-to-flux convertor, if we use a nominal flux [13] $\Phi_{\theta j}$ to redefine $\theta_j$, namely

$$\Phi_{\theta j} = \frac{\Phi_0}{2\pi}\theta_j; j=(1,\ldots,Q) \qquad (2)$$

where $\Phi_0 = 2.07 \times 10^{-15}$ Wb.

Base on this nominal flux concept, the current-to-flux functions of those AJs are expressed in matrix as

$$F_J(\mathbf{\Phi_\theta},\mathbf{i_\theta},\mathbf{I_b})=0 \Leftrightarrow$$
$$\mathbf{i_\theta} = \mathbf{C}\cdot\frac{d^2\mathbf{\Phi_\theta}}{dt^2} + \mathbf{G}\cdot\frac{d\mathbf{\Phi_\theta}}{dt} + \left(\mathbf{I_0}\cdot\sin\frac{2\pi\mathbf{\Phi_\theta}}{\Phi_0} - \mathbf{I_b}\right) \qquad (3)$$

where $\mathbf{\Phi_\theta} = [\Phi_{\theta 1}, \ldots, \Phi_{\theta Q}]^T$, $\mathbf{i_\theta} = [i_{\theta 1}, \ldots, i_{\theta Q}]^T$, and $\mathbf{I_b} = [I_{b1}, \ldots, I_{bQ}]^T$, are the state vectors of AJs; $\mathbf{C} = diag\{C_1, \ldots, C_Q\}$, $\mathbf{G} = diag\{1/R_1, \ldots, 1/R_Q\}$, and $\mathbf{I_0} = diag\{I_{01}, \ldots, I_{0Q}\}$, are the diagonal matrices of circuit parameters.

Therefore, loops and AJs are the only two objects of Josephson junction circuits. Their inputs and outputs are associated with two circuit laws.

First, the outputs $\mathbf{\Phi_\theta}$ of the AJs and the fluxes $\mathbf{\Phi_m}$ coupled in the loops comply with the flux-quantization law (FQL), namely,

$$\mathbf{\Phi_m} + \mathbf{\sigma}\cdot\mathbf{\Phi_\theta} = \mathbf{n}\cdot\Phi_0 \qquad (4)$$

where $\mathbf{n} = [n_1, \ldots, n_P]^T$, and $n_i$ ($i = 1, 2, \ldots, Q$) is an integer used to define the number of flux quanta trapped inside Loop-$i$; $\mathbf{\sigma}$ is the loop-to-junction incidence matrix that describes the relation between AJs and loops, as illustrated in Fig. 3(d).

Second, the relation between $\mathbf{i_m}$ and $\mathbf{i_\theta}$ is derived with the Kirchhoff's current law (KCL) as

$$\mathbf{i_\theta} = \mathbf{\sigma}^T \cdot \mathbf{i_m} \qquad (5)$$

where, $\mathbf{\sigma}^T$ is the transpose of $\mathbf{\sigma}$.

*B. Behaviors of Loops and AJs*

A system model of Josephson junction circuits is drawn by synthetizing the circuit equations from (1) to (5), as shown in Fig. 5. This system model clearly depicts the behaviors of AJs and loops:

First, AJs with the current-to-flux functions work as the magnetic-flux pumps that generate flux outputs $\mathbf{\Phi_\theta}$ into loops. The behavior of AJ-$j$ is analogous to a classical particle rolling in a washboard potential [13] $U_j$, as shown in Fig. 6. This potential $U_j$ is defined as

$$U_j = \int_0^{\Phi_{\theta j}} (I_{0j}\sin\frac{2\pi\Phi_{\theta j}}{\Phi_0} - I_{bj}) \cdot d\Phi_{\theta j}$$
$$= \frac{\Phi_0 I_{0j}}{2\pi}(1-\cos\frac{2\pi\Phi_{\theta j}}{\Phi_0}) - I_{bj}\Phi_{\theta j}; (j=1,\cdots,Q) \qquad (6)$$

It consists of a linear part and a cosine part. The slope of the linear part is set by the bias current $I_{bj}$, and the amplitude of the cosine part is decided by the critical current $I_{0j}$. In Fig. 6, Case-I with $I_{bj} = 0.3I_{0j}$ depicts the potential usually set for the AJs of SFQ circuits; Case-II with $I_{bj} = 1.1 I_{0j}$ shows the one set for the AJs of SQUID circuits.

In those potentials, AJ-$j$ is pushed by its branch current $i_{\theta j}$. It stays still in the floors of the potential valley, if $i_{\theta j} < I_{THj}$, and will step forward, as long as $i_{\theta j} > I_{THj}$. This $I_{THj}$ can be derived by finding the maximum $dU_j/d\Phi_{\theta j}$ from (6) as

$$I_{THj} \equiv I_{0j} - I_{bj} \qquad (7)$$

Therefore, AJ-$j$ is turned on and off according to a threshold current $I_{THj}$, like a diode. The input vector $\mathbf{I_b}$ in Fig. 5 is actually used to adjust the threshold currents of AJs.

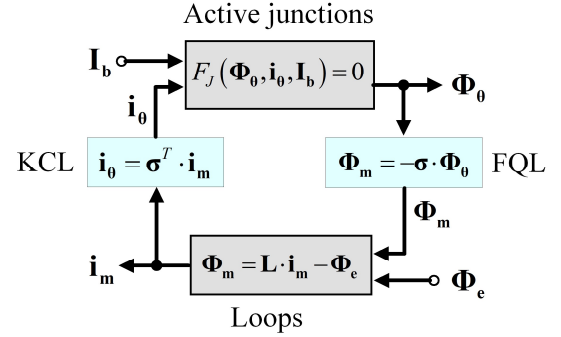

Fig. 5. System model of Josephson junction circuits, where AJs execute the current-to-flux functions, while loops implement the flux-to-current conversions. two transfer functions are associated with KCL and FQL.

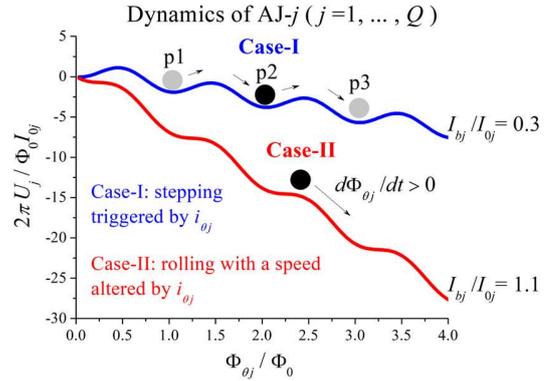

Fig. 6. Particle analogy of AJ-$j$: in Case-I with $I_{bj} = 0.3\ I_{0j}$, the particle stays at the floor of potential valley, and steps forward one flux-quantum each time when it is triggered by $i_{\theta j}$; in Case-II with $I_{bj} = 1.1\ I_{0j}$, the particle will keep rolling downhill with a speed altered by $i_{\theta j}$.

Second, superconducting loops are the magnetic-flux containers; they implement the linear flux-to-current functions to turn $\mathbf{\Phi_e}$ and $\mathbf{\Phi_m}$ into loop currents $\mathbf{i_m}$ flowing through AJs. The input $\mathbf{\Phi_e}$ in Fig. 5 describes both the trapped flux and external flux of loops, after it is combined the $\mathbf{n}\cdot\Phi_0$ in (4).

According to (4), the fluxes $\mathbf{\Phi_m}$ contained in loops are contributed by outputs of AJs. They are also quantized with $\Phi_0$, because the $\Phi_{\theta j}$ of AJ-$j$ is always located near $n\Phi_0$ ($n$ is an integer) when AJ-$j$ is switched off, as illustrated in Case-I. Correspondingly, superconducting loops can be viewed as the flux-quantum containers; they are either empty or occupied with one or several flux quanta.

*C. MFF Diagram*

The MFF diagram of a Josephson junction circuit is a kind of interaction diagram of loops and AJs. The symbols and connections for drawing MFF diagrams are illustrated in Fig. 7, where Loop-$i$ is abstracted as a circle and AJ-$j$ is symbolized with a brick-shape block, as shown in Fig. 7(a) and (c). The Outer-loop [22] with a loop-current fixed as zero is represented with a bar shown in Fig. 7(b); it is a special loop used to terminate the flux-flow of AJs, similar to the 'ground' in conventional circuit diagrams.

Meanwhile, the elements of **σ** and **L_m** are implemented with two types of directed lines connected between loops and AJs, as illustrated in Fig. 7(d). For example, $\sigma_{ij} = 1$ is represented with a directed line connected from Loop-$i$ to AJ-$j$; $\sigma_{ik} = -1$ is exhibited with a line connected from AJ-$j$ to Loop-$k$; a mutual inductance $M_{ik}$ is expressed a double-arrow line connected between Loop-$i$ and Loop-$k$.

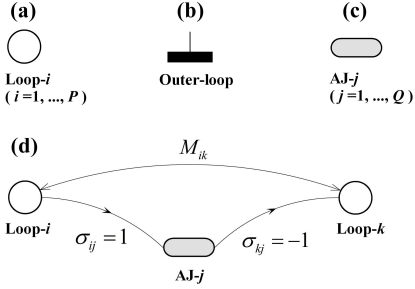

Fig. 7. Symbols and connections of MFF diagrams: (a) symbol of loops; (b) symbol of the outer-loop; (c) symbol of AJs; (d) two types of connections between loops and AJs.

A MFF diagram is the graphical implementation of the system model shown in Fig. 5. It intuitively depicts how AJs transfer the magnetic fluxes from loop to loop. For example, in Fig. 7(d), Loop-$i$ with a self-inductance $L_i$ is a magnetic-flux container; it turns $\Phi_{mi}$ and $\Phi_{ei}$, as well as the flux $M_{ik}i_{mk}$ generated by Loop-$k$ into loop current $i_{mi}$; AJ-$j$ is a diode-like magnetic-flux pump driven by the branch current $i_{\theta j}$; it is pushed forward by $i_{mi}$ of Loop-$i$, and hindered by $i_{mk}$ of Loop-$k$, for $\sigma_{ij} = 1$ and $\sigma_{kj} = -1$. When $i_{\theta j} > I_{THj}$, AJ-$j$ will achieve one flux-quantum increasement in $\Phi_{\theta j}$. Consequently, one flux-quantum will be eliminated from Loop-$i$ and added to Loop-$k$. It is appeared that AJ-$j$ pumps a flux-quantum from Loop-$i$ to Loop-$k$ along the directed lines.

Therefore, based on the behaviors of AJs and loops, we can use MFF diagrams to trace the variations of flux quanta contained in loops and visualize the flux-quantum flows between loops. MFF diagrams intuitively depict the dynamics of flux quanta transferring inside Josephson junction circuits.

The MFF diagrams of various Josephson junction circuits are all composed of basic MFF components. The detailed working principles of those typical MFF components are demonstrated in Appendix.

### III. APPLICATION EXAMPLE

*A. Dc-SQUID*

The dc-SQUID circuit shown in Fig. 1(a) is redrawn with AJs, as shown in Fig. 8 (a). Its MFF diagram consists of a flux-quantum generator with $I_{TH1} < 0$ and a flux-quantum absorber with $I_{TH2} < 0$, as shown in Fig. 8(b), where a small $i_{m1}$ will first trigger the flux-quantum generator to pump flux quanta into Loop-1 and increase $i_{m1}$, while the increased $i_{m1}$ will trigger the flux-quantum absorber to pump the flux quanta out and decrease $i_{m1}$.

The MFF diagram reveals that the dc-SQUID is a simple system with flux quanta pumped in and out repeatedly; the voltage measured at the dc-SQUID is actually the average flow-rate of AJ-1 or AJ-2; it is modulated by the external flux $\Phi_{in}$, because $\Phi_{in}$ varies the $i_{m1}$ and modifies the trigger timing of two AJs.

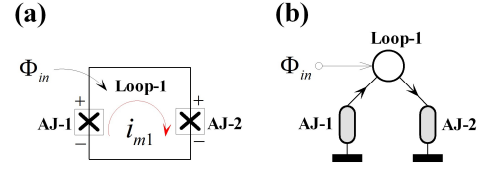

Fig. 8. (a) Two AJs and one loop defined in the dc-SQUID circuit. (b) MFF diagram of the dc-SQUID.

*B. TTL-to-SFQ*

The dc-SQUID with $I_{TH1} > 0$ and $I_{TH2} > 0$ works as a TTL-to-SFQ convertor used to transform the TTL logics to SFQ logics, as shown in Fig. 9(a), where the AJ-1 and AJ-2 can only be turned on by the $\Phi_{in}$ generated by a TTL signal. Its working principle can be read from the MFF diagram shown in Fig. 9(b). When the TTL signal outputs a high-level voltage, it generates a negative $\Phi_{in}$ and induces a negative $i_{m1}$ in Loop-1; the negative $i_{m1}$ will turn on the AJ-1 to pump a flux-quantum into Loop-1. When the TTL signal returns to the low-level, it induces a positive $i_{m1}$ and turns on the AJ-2 to pump a flux-quantum out of Loop-1.

From the MFF diagram, we can clearly find that the '0' and '1' states of TTL logics are transformed into the 'empty' and 'occupied' states of circuit loops for SFQ logics.

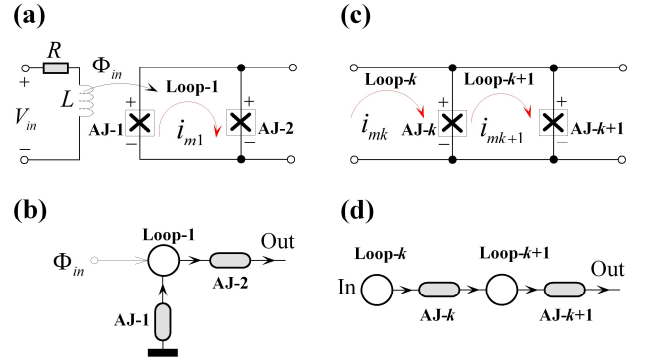

Fig. 9. (a) Conventional circuit diagram of the TLL-to-SFQ circuit redrawn with AJs, and (b) the corresponding MFF diagram. (c) Circuit diagram redrawn with AJs and (d) the MFF diagram for the JTL circuit.

*C. Josephson Transmission Line*

The cell circuit of Josephson transmission line (JTL) shown in Fig. 1(b) is redrawn with its AJs, as shown in Fig. 9(c). Its MFF diagram shown in Fig. 9(d) exhibits that the JTL is the flux-quantum pipelines connected in series, where AJ-$k$ transfers flux-quanta from Loop-$k$ to Loop-$k$+1 one by one, like a relay.

*D. D-type Flip-Flop Circuit*

The D-type flip-flop (DFF) circuit shown in Fig. 1(c) is redrawn with four AJs, as shown in Fig. 10(a). Its MFF diagram is shown in Fig. 10(b), where AJ-2, AJ-3 and Loop-2 constitute an unbuffered flux-quantum pipeline; Loop-1, Loop-3 and AJ-4 compose a flux-quantum merger. From this MFF diagram, we can clearly see how the general clock synchronize the input signal; a flux-quantum input in Loop-2 will be first transferred to Loop-3 by an unbuffered pipeline, then it will be stored in Loop-3; finally, it is merged with the flux-quantum in Loop-1 filled by the clock signal.

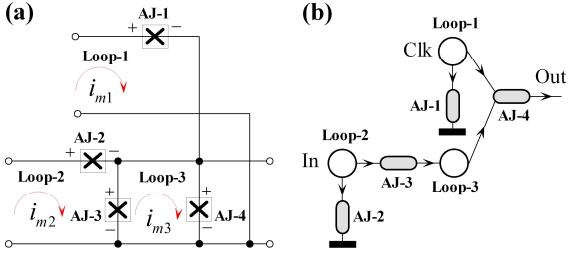

Fig. 10. (a) Conventional circuit diagram of the DFF circuit redrawn with four AJs. (b) MFF diagram of the DFF circuit.

### E. Synchronous AND-gate Circuit

The MFF diagram of the AND-gate circuit shown in Fig. 1(d) is drawn as shown in Fig. 11. The input flux quanta in Loop-2 and Loop-4 are firstly transferred to Loop-7 and Loop-8, through two DFFs synchronized by the clock signals in Loop-1 and Loop-3; they are then merged by an AND-type merger (shown in Fig. 17) with Loop-7 and Loop-8 as input loops.

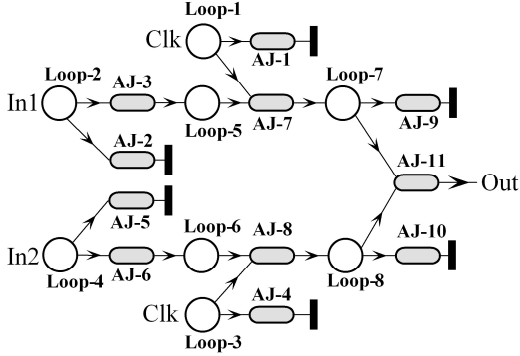

Fig. 11. MFF diagram of the synchronous AND-gate.

### F. Synchronous OR-gate

The MFF diagram of the OR-gate circuit shown in Fig. 1(e) is exhibited in Fig. 12. The flux-quantum in either Loop-2 or Loop-3 will be firstly transferred to Loop-6 by an OR-type merger (illustrated in Fig. 18) with Loop-4 and Loop-5 as input loops, to realize the OR-logic; the flux-quantum in Loop-6 will be then ejected out by AJ-8, under the synchronization of the clock flux-quantum pumped into Loop-1.

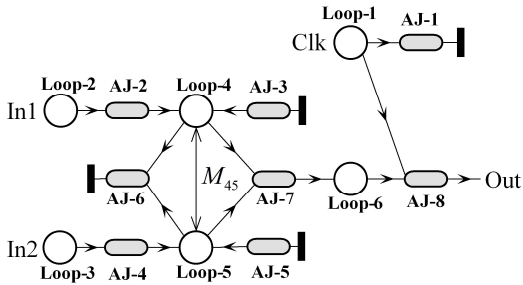

Fig. 12. MFF diagram of the synchronous OR-gate.

### G. Synchronous NOT-gate

The MFF diagram of the NOT-gate circuit shown in Fig. 1(f) is depicted in Fig. 13. It is exactly the flux-quantum multiplexer demonstrated in Fig. 19, which logic is that the flux-quantum in Loop-1 will not flow out through AJ-2, if Loop-2 is already occupied before Loop-1 receives a new flux-quantum.

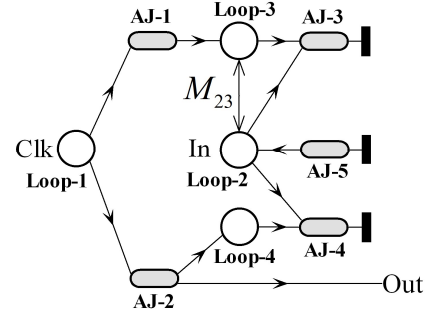

Fig. 13. MFF diagram of the synchronous NOT-gate.

## IV. DISCUSSION AND CONCLUSION

We developed a type of MFF diagrams and the corresponding flux-based analysis method to expand the circuit analysis methodologies for the magnetic flux dominant electric circuits, and demonstrated their applications in the analyses of typical Josephson junction circuits. The advantages of MFF diagrams are summarized as follows.

(1) MFF diagrams make Josephson junction circuits to be understood as easily as CMOS circuits. In a MFF diagram of Josephson junction circuits, the dynamics of magnetic flues transferring from loop to loop are visualized; all the circuit components and circuit variables achieves new physical meanings. For instance, superconducting loops are the flux-quantum containers; their coupled magnetic fluxes are the logic bits for SFQ circuits; Josephson junctions are the diode-like magnetic-flux pumps; their voltage pulses at their two terminals depict the flux flow-rate through the pumps.

(2) MFF diagrams simply the design and analysis of Josephson junction circuits significantly. The MFF diagram of an electric circuit is an object diagram of AJs and loops to implement the system model shown in Fig. 5. After a MFF diagram is designed, it can be directly simulated with the general circuit equations for verification, and converted into the conventional circuit diagram for physical implementation as well, according the definitions in Fig. 3.

(3) MFF diagrams bridge the gap between superconducting Josephson junction circuits and normal RLC circuits. In Fig. 7, Loop-$i$ represents a non-superconducting loop, if the $n_i$ is set as a real number; AJ-$j$ becomes a normal component of resistor and capacitor in parallel, if $I_{01} = 0$. MFF diagrams unify the design and analysis for both superconducting and normal electric circuits, as well as their hybrids.

The MFF diagram and conventional circuit diagram of an electric circuit are two pictures of the circuit viewed from the magnetic and electric fields, respectively. They are complementary to each other in the design and analysis of the magnetic flux sensitive electric circuits, such as the Josephson junction circuits.

APPENDIX

## A. Flux-Quantum Generator and Absorber

*Flux-quantum generator*: it is implemented with an AJ connected between a loop and the outer-loop, as illustrated in Fig. 14(a), where AJ-1 will pump one flux-quantum to the empty Loop-1, if it is turned on by a negative $i_{m1}$. The negative $i_{m1}$ can be induced by a negative $\Phi_{e1}$, or the fluxes of other loops through mutual inductances.

*Flux-quantum absorber*: it is used to dump the flux-quantum in Loop-1 to the outer-loop, as illustrated in Fig. 14(b), which is dual to the flux-quantum generator.

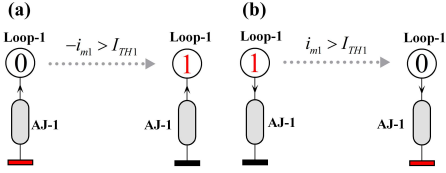

Fig. 14. (a) Flux-quantum generator, where '0' and '1' mark the number of the flux quanta contained in Loop-1 to make the state of loops visible; AJ-1 is connected between Loop-1 and the outer-loop, with a directed line pointing to Loop-1; it will transfer one flux-quantum to Loop-1, when $-i_{m1} > I_{TH1}$; $i_{m1}$ in Loop-1 can be induced by a negative $\Phi_{e1}$. (b) Flux-quantum absorber, where AJ-1 is connected with Loop-1 by a directed line pointing to AJ-1; it will absorb one flux-quantum from Loop-1, when $i_{m1} > I_{TH1}$.

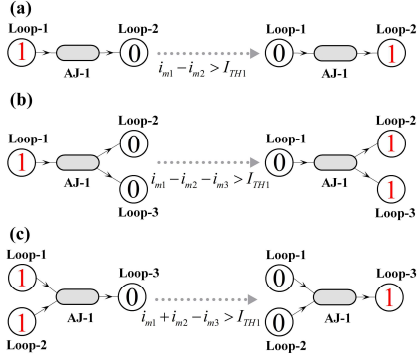

Fig. 15. (a) Flux-quantum pipeline. AJ-1 is connected between Loop-1 and Loop-2, with $i_{\theta1}=i_{m1} - i_{m2}$; the flux-quantum in Loop-1 flows to Loop-2, when $i_{\theta1} > I_{TH1}$. (b) Flux-quantum splitter, where AJ-1 transfers flux quanta from Loop-1 to Loop-2 and Loop-3, and $i_{\theta1}=i_{m1} - i_{m2} - i_{m3}$. AJ-1 will absorb one flux-quantum from Loop-1 and fill one flux-quantum to both Loop-2 and Loop-3, when $i_{\theta1} > I_{TH1}$. (c) Flux-quantum merger, where $i_{\theta1}=i_{m1} + i_{m2} - i_{m3}$. AJ-1 eliminates one flux-quantum in both Loop-1 and Loop-2, and fills one flux-quantum into Loop-3, when $i_{\theta1} > I_{TH1}$.

## B. Flux-Quantum Transmitters

There are four basic flux-quantum transmitters in Josephson junction circuits.

*Flux-quantum pipeline*: it is used to transfer flux quanta from one loop to another, as illustrated in Fig. 15(a), where AJ-1 will transfer one flux-quantum from Loop-1 to Loop-2, when $i_{\theta1} > I_{TH1}$.

*Flux-quantum splitter*: it is used to split the input flux-quantum to two loops, as depicted in Fig. 15(b), where AJ-1 duplicates the flux-quantum received by Loop-1 and assigns them to Loop-2 and Loop-3.

*Flux-quantum merger*: it merges two input flux quanta into one and pumps it to the next loop, as illustrated in Fig. 15(c), in which AJ-1 absorbs two flux quanta from Loop-1 and Loop-2, and fills one flux-quantum into Loop-3.

*Unbuffered Flux-quantum pipeline*: it is the combination of a flux-quantum pipeline and a flux-quantum absorber, as shown in Fig. 16. One flux-quantum in Loop-1 will be transferred to Loop-2 by AJ-1, if Loop-2 is empty, as illustrated in Fig. 16(a); it will be dumped by AJ-2, if Loop-2 has been already occupied, as illustrated in Fig. 16(b).

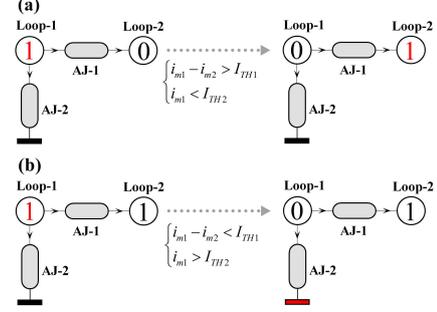

Fig. 16. Unbuffered flux-quantum pipeline, which is a flux-quantum pipeline with a flux-quantum absorber combined at the input loop: (a) the flux-quantum in Loop-1 will be transferred to loop-2 by AJ-1, if loop-2 is empty; (b) it will be dumped by AJ-2, if Loop-2 is already occupied.

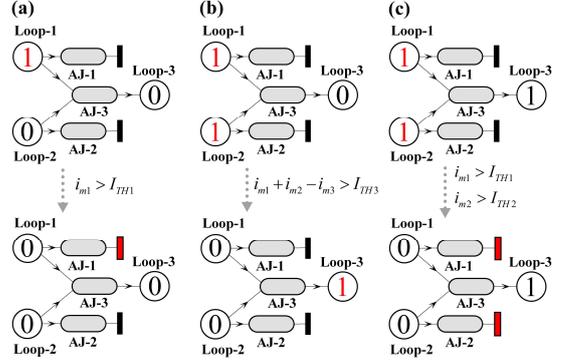

Fig. 17. AND-type merger. It is a flux-quantum merger with two flux-quantum absorbers configured at input loops: (a) when loop-3 and loop-2 are empty, the flux-quantum contained in loop-1 will be absorbed by AJ-1; (b) when loop-3 is empty, the flux quanta in both Loop-1 and loop-2 will be merged and transferred to loop-3; (c) when loop-3 is occupied, the flux quanta in both loop-1 and loop-2 will be dumped through the flux-quantum absorbers.

## C. Basic Logic Circuits

There are three basic logic components for SFQ logic gates.

*AND-type merger*: it is a merger with two additional flux-quantum absorbers configured at two input loops, as depicted in Fig. 17. It implements the AND-logic in transferring the flux quanta in Loop-1 and Loop-2. For instance, only when Loop-1 and Loop-2 receive one flux-quantum at the same time will AJ-3 fill one flux-quantum to the empty Loop-3, as illustrated in Fig. 17(b); if one input loop is empty or Loop-3 is occupied, the flux-quantum in input loops will be dumped by the flux-quantum absorbers, as shown in Fig. 17(a) and (c).

*OR-type merger*: it merges the flux quanta in two input loops according to the OR-logic, as demonstrated in Fig. 18, where each input loop is configured with a flux-quantum generator; two input loops are coupled with a negative $M_{12}$ and are connected to a flux-quantum absorber implemented by AJ-1. With the aid of $M_{12}$, a flux-quantum in either input loop will induce a flux-quantum to another input loop; Two flux quanta will be merged into the empty Loop-3 by AJ-2, as illustrated in Fig. 18(a), or be absorbed by AJ-1 if Loop-3 is already occupied, as shown in Fig. 18(b).

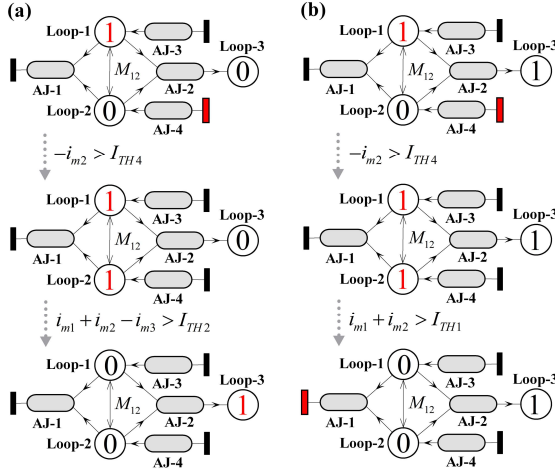

Fig. 18. OR-type merger. It is a flux-quantum merger with two flux-quantum generators connected at two input loops; two input loops are connected to a flux-quantum absorber implemented by AJ-1, and are coupled through a negative $M_{12}$; with the aid of $M_{12}$, one flux-quantum received by any one input loop will induce a flux-quantum to another input loop: (a) two flux quanta are merged and transferred to loop-3, if Loop-3 is empty; (b) they will be absorbed by AJ-1, if Loop-3 is already occupied.

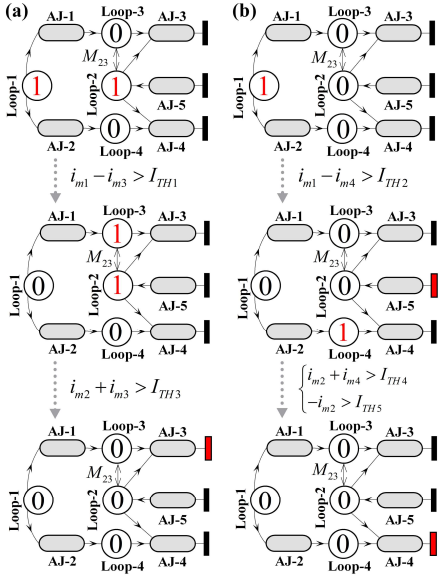

Fig. 19. Flux-quantum multiplexer, where Loop-2 and Loop3 are coupled through a mutual-inductance $M_{12}$. It transfers the flux-quantum in Loop-1 to either Loop-3 or Loop-4, according to the state of Loop-2: (a) if Loop-2 is already occupied before Loop-1 receives a flux-quantum, the flux-quantum in Loop-1 will be transferred to Loop-3 by AJ-1, because AJ-1 is turned on first by a smaller $i_{m3}$ of Loop-3, which is induced by the flux-quantum in Loop-2 through $M_{12}$. (b) if Loop-2 is empty before Loop-1 receives a flux-quantum, the flux-quantum in Loop-1 will be transferred to Loop-4 by AJ-2, and is finally absorbed by AJ-4, accompanied with the flux-quantum generated by AJ-5 in Loop-2.

*Flux-quantum multiplexer*: it transfers the flux-quantum in one input loop through two paths selected by the flux-quantum in another input loop, as illustrated in Fig. 19, where $M_{23}$ is a negative mutual-inductance between Loop-2 and Loop-3. The flux-quantum in Loop-1 will flow into Loop-3, if Loop-2 is occupied, as illustrated in Fig. 19(a); it will flow into Loop-4, if Loop-2 is empty, as demonstrated in Fig. 19(b).